\documentclass[a4paper,prb,twocolumn,showpacs,amsmath,amssymb]{revtex4}
\usepackage[english]{babel}
\usepackage[latin1]{inputenc}
\usepackage{amscd,amsmath,amssymb,verbatim}
\usepackage[dvips]{graphicx,color}
\usepackage{subfigure}
\usepackage{textcomp}
\usepackage{calc}
\usepackage{bbm}
\usepackage[OT1,T1]{fontenc}
\usepackage{natbib}
\usepackage[verbose,hypertexnames=false,bookmarksopenlevel=1,filecolor=blue,
linkcolor=blue,citecolor=blue,pdfstartview=FitH,bookmarksopen,bookmarksnumbered,
colorlinks,plainpages=false,linktocpage,ps2pdf]{hyperref}
\usepackage[]{psfrag}

\DeclareMathAlphabet{\mathpzc}{OT1}{pzc}{m}{it}

\DeclareSymbolFont{mathbold}{OML}{cmm}{b}{it}
\DeclareMathSymbol{\balpha}{\mathord}{mathbold}{11}
\DeclareMathSymbol{\bbeta}{\mathord}{mathbold}{12}
\DeclareMathSymbol{\bgamma}{\mathord}{mathbold}{13}
\DeclareMathSymbol{\bdelta}{\mathord}{mathbold}{14}
\DeclareMathSymbol{\bepsilon}{\mathord}{mathbold}{15}
\DeclareMathSymbol{\bvarepsilon}{\mathord}{mathbold}{34}
\DeclareMathSymbol{\bzeta}{\mathord}{mathbold}{16}
\DeclareMathSymbol{\bEta}{\mathord}{mathbold}{17}
\DeclareMathSymbol{\btheta}{\mathord}{mathbold}{18}
\DeclareMathSymbol{\bvartheta}{\mathord}{mathbold}{35}
\DeclareMathSymbol{\biota}{\mathord}{mathbold}{19}
\DeclareMathSymbol{\bkappa}{\mathord}{mathbold}{20}
\DeclareMathSymbol{\blambda}{\mathord}{mathbold}{21}
\DeclareMathSymbol{\bmu}{\mathord}{mathbold}{22}
\DeclareMathSymbol{\bnu}{\mathord}{mathbold}{23}
\DeclareMathSymbol{\bxi}{\mathord}{mathbold}{24}
\DeclareMathSymbol{\bpi}{\mathord}{mathbold}{25}
\DeclareMathSymbol{\bvarpi}{\mathord}{mathbold}{36}
\DeclareMathSymbol{\brho}{\mathord}{mathbold}{26}
\DeclareMathSymbol{\bvarrho}{\mathord}{mathbold}{37}
\DeclareMathSymbol{\bsigma}{\mathord}{mathbold}{27}
\DeclareMathSymbol{\bvarsigma}{\mathord}{mathbold}{38}
\DeclareMathSymbol{\btau}{\mathord}{mathbold}{28}
\DeclareMathSymbol{\bupsilon}{\mathord}{mathbold}{29}
\DeclareMathSymbol{\bphi}{\mathord}{mathbold}{30}
\DeclareMathSymbol{\bvarphi}{\mathord}{mathbold}{39}
\DeclareMathSymbol{\bchi}{\mathord}{mathbold}{31}
\DeclareMathSymbol{\bpsi}{\mathord}{mathbold}{32}
\DeclareMathSymbol{\bomega}{\mathord}{mathbold}{33}
\DeclareMathSymbol{\biGamma}{\mathord}{mathbold}{0}
\DeclareMathSymbol{\biDelta}{\mathord}{mathbold}{1}
\DeclareMathSymbol{\biTheta}{\mathord}{mathbold}{2}
\DeclareMathSymbol{\biLambda}{\mathord}{mathbold}{3}
\DeclareMathSymbol{\biXi}{\mathord}{mathbold}{4}
\DeclareMathSymbol{\biPi}{\mathord}{mathbold}{5}
\DeclareMathSymbol{\biSigma}{\mathord}{mathbold}{6}
\DeclareMathSymbol{\biUpsilon}{\mathord}{mathbold}{7}
\DeclareMathSymbol{\biPhi}{\mathord}{mathbold}{8}
\DeclareMathSymbol{\biPsi}{\mathord}{mathbold}{9}
\DeclareMathSymbol{\biOmega}{\mathord}{mathbold}{10}

\newcommand{\eins}{\mathbbm{1}}

\newcommand{\ket}[1]{|#1\rangle}

\newcommand{\kt}[1]{\text{\tiny{#1}}}

\newcommand{\I}{\mathit{i}}

\newcommand{\sinc}{\sin\!\text{c}}

\begin{document}
\title{Direction Dependence of Spin Relaxation in Confined 2D Systems}

\author{P. Wenk}
\email[]{p.wenk@jacobs-university.de (corresponding author)}
\homepage[]{www.physnet.uni-hamburg.de/hp/pwenk/}
\affiliation{School of Engineering and Science, Jacobs University Bremen, Bremen 28759, Germany}
\author{S. Kettemann}
\email[]{s.kettemann@jacobs-university.de}
\homepage[]{www.jacobs-university.de/ses/skettemann}
\affiliation{School of Engineering and Science, Jacobs University Bremen, Bremen 28759, Germany, and Asia Pacific Center for Theoretical
Physics and Division of Advanced Materials Science Pohang University of Science and Technology (POSTECH) San31, Hyoja-dong, Nam-gu, Pohang 790-784, South Korea}

\begin{abstract}
The dependence of spin relaxation on the direction of the quantum wire under Rashba and Dresselhaus (linear and cubic) spin orbit coupling is studied. Comprising the dimensional reduction of the wire in the diffusive regime, the lowest spin relaxation and dephasing rates for (001) and (110) systems are found. The analysis of spin relaxation reduction is then extended to non-diffusive wires where it is shown that, in contrast to the theory of dimensional crossover from weak localization to weak antilocalization in diffusive wires, the relaxation due to cubic Dresselhaus spin orbit coupling is reduced and the linear part shifted with the number of transverse channels.
\end{abstract}
\maketitle
\section{Introduction}
Spin dynamics in semiconductors have been studied for decades, but still the prime condition for building spintronic devices, namely the understanding of spin relaxation, is not satisfactorily fulfilled. In the following we focus on materials where the dominant mechanism for spin relaxation is governed by the D'yakonov-Perel spin relaxation (DPR).\cite{perel} This mechanism results from lifting the spin degeneracy which is due to time inversion symmetry and spacial inversion symmetry and leads to the effect of slower spin dephasing the faster the momentum relaxes (motional narrowing\cite{Bloembergen1948,Slichter1989,wenkbook}). In one of the most studied systems GaAs/AlGaAs DPR is the most relevant mechanism in the metallic regime.\cite{dzhioev}\\
Preserving time reversion symmetry, the spin splitting can be due to bulk inversion asymmetry (BIA)\cite{dresselhaus} and also due to the asymmetry arising from the structure of the quantum well (QW), the structure inversion asymmetry (SIA)\cite{rashba_1}.
In Ref.\,\onlinecite{Kettemann:PRL98:2007,Wenk2010} it was shown how the spin relaxes in a quasi 1D electron system in a QW grown in the $[001]$ direction, depending on the width of the wire, where the normal of the boundary was pointing in the $[010]$ direction. It is already known that in a (001) 2D system with BIA and SIA we get an anisotropic spin-relaxation.\cite{Kainz2003,Cheng2007,Wu2010} This has also been studied numerically in quasi-1D GaAs wires\cite{Liu2009}. In this work, Sec.\,\ref{sec:001grothDir}, we present analytical results concerning this anisotropy for the 2D case as well as the case of QW with spin and charge conserving boundaries.\\
We also extend our analysis to other growth directions, Sec.\,\ref{sec:110system}: Searching for long spin decoherence times at room temperature, the (110) QW attracted attention.\cite{Adachi200136,Dohrmann2004} The properties of spin relaxation in systems with this growth direction has also been related to weak localization (WL) measurements, Ref.\,\onlinecite{Hassenkam1997}. We present analytical explanations for dimensional spin relaxation reduction and discuss the crossover from WL to weak antilocalization (WAL), Sec.\,\ref{sec:WL}.\\
As we will show in the following sections, the cubic Dresselhaus spin orbit coupling (SOC) gives always rise to a limitation of the spin relaxation time in the diffusive case, $W\gg l_e$, with the wire width W and the elastic mean free path $l_e$. However some of the experiments are done on ballistic wires and we need to modify the theory used in Ref.\,\onlinecite{Kettemann:PRL98:2007,Wenk2010} to enable us to study the crossover from diffusive to ballistic wires. In Sec.\,\ref{sec:DiffBallis} we show how the spin relaxation which is due to cubic Dresselhaus SOC reduces with the number of channels in the QW.
\\
We consider the following Hamiltonian with SOC
\begin{equation} \label{hamiltonian}
H = \frac{1}{2 m_e}   ({\bf p} +e {\bf A} )^2 +V({\bf x})
-\frac{1}{2} \gamma\bsigma\left({\bf B}+{\bf B}_\kt{SO}({\bf p})\right),
\end{equation}
where  $m_e$ is the effective electron mass.  
 ${\bf A}$ is the vector potential due to the external magnetic field 
  ${\bf B}$.  ${\bf B}_\kt{SO}^T = ({B_\kt{SO}}_x, {B_\kt{SO}}_y) $ is the momentum dependent SO field. $\bsigma$ is a vector, with components $\bsigma_i$, $i =x,y,z$,  the Pauli matrices, $\gamma$ is the gyromagnetic ratio with $\gamma=g \mu_\kt{B}$ with the effective g factor of the material, and $\mu_\kt{B} = e/2m_e$ is the Bohr magneton constant. For example, III-V and II-VI semiconductors such as GaAs, InSb have zinc-blend structure. This BIA causes a SO interaction, which, to lowest order in the wave vector $\bf k$, is given by
\cite{dresselhaus}
\begin{equation}
-\frac{1}{2}\gamma {\bf B}_\kt{SO,D}= \gamma_D\sum_i\hat{e}_i p_i(p_{i+1}^2-p_{i+2}^2)\label{Dresselhaus}
\end{equation}
where the principal crystal axes are given by $i\in\{x,y,z\},i\rightarrow ((i-1) \mod 3)+1$ and the spin-orbit coefficient for the bulk semiconductor $\gamma_D$. We consider the standard white-noise model for  the impurity potential, $V({\bf x})$, which vanishes on average
$\langle V ({\bf x}) \rangle=0$, is uncorrelated, $\langle V({\bf
x}) V({\bf x'})\rangle = \delta ({\bf x-x'})/2 \pi \nu \tau$, and
weak, $\epsilon_F\tau\gg 1$. Here, $\nu = m_e/(2 \pi)$ is the
average density of states per spin channel, $\epsilon_F$ is the Fermi energy and $\tau$ is the elastic
scattering time. To address both, the WL corrections as well as the spin relaxation rates in the system, we analyze the Cooperon\cite{HIKAMI1980}
\begin{align}\label{Cooperon0}
{}&\hat{C}({\bf Q})^{-1}=  \frac{1}{\tau} \left( 1 -
\int \frac{d \varphi}{2\pi} \right. \nonumber\\
{}&\left.\frac{1}{1+\I\tau(\mathbf{v}(\mathbf{Q}+2e\mathbf{A}+2m_e\hat a \mathbf{S}) + H_{\sigma'}
+ H_Z  )}\right),
\end{align}
where the integral is performed over all angles of velocity $\mathbf v$ on the Fermi surface, $ H_{\sigma'}=-(\mathbf{Q}+2 e\mathbf{A})\hat{a} \bsigma^\prime$ and the Zeeman coupling to the external magnetic field yields,
\begin{equation}
H_Z=- \frac{1}{2} \gamma (\bsigma^\prime-\bsigma)\mathbf{B}.\label{zeemanTerm}
\end{equation}
The coupling between the orbital motion and the spin ${\mathbf S=(\bsigma+\bsigma^\prime)/2}$ is described by the SOC operator $\hat a$. The spin quantum number is 1 instead of $1/2$ due to the electron-hole excitation.
It follows that for weak disorder and without Zeeman coupling, the Cooperon depends only on the total momentum $\mathbf{Q}$ and the total spin $ {\bf S}$. Expanding the Cooperon to second order in $( {\bf Q} +2 e {\bf A} + 2m_e \hat{a} {\bf S} )$ and performing  the angular integral which  is for 2D diffusion (elastic mean free path $l_e$  smaller than wire width $W$) continuous from $0$ to $2 \pi$, yields:
\begin{equation}
  \hat{C} ({\bf Q})= \frac{1}{D_e( {\bf Q} + 2 e {\bf A} + 2 e  {\bf A}_{\bf S})^2 + H_{\gamma_D}  }.\label{Cooperon1}
\end{equation}
The effective vector potential due to SO interaction is ${\bf A}_{\bf S} = m_e  \hat{\alpha} {\bf S}/e$, where $\hat{\alpha} =
\langle\hat{a}\rangle$ is averaged over angle. The SO term $H_{\gamma_D}$, which cannot be rewritten as a vector potential, is in our case due to the appearance of cubic Dresselhaus SOC.

\subsection{Example}
To get an idea of the procedure we recall the situation presented in Ref.\,\onlinecite{Kettemann:PRL98:2007,Wenk2010}. Spin relaxation in a (001) quasi-1D wire in [100] direction:\\
The Dresselhaus term, Eq.\,(\ref{Dresselhaus}), for QWs grown in the $[001]$ direction is given by
\cite{dresselhaus}
\begin{equation}
-\frac{1}{2} \gamma{\bf B}_\kt{SO,D} = \alpha_1 ( -\hat{e}_x p_x + \hat{e}_y p_y)
+ \gamma_D  ( \hat{e}_x p_x p_y^2 - \hat{e}_y p_y p_x^2).
\end{equation}
Here, $\alpha_1 = \gamma_D \langle p_z^2 \rangle$ is the linear Dresselhaus parameter, which measures the strength of the term  linear in    momenta $p_x, p_y$ in the plane of the 2D electron system (2DES). When  $\langle p_z^2 \rangle \sim 1/a^2 \ge k_F^2$  ($a$ is the  thickness of the 2DES, $k_F$, Fermi wave number), that term  exceeds the cubic Dresselhaus terms which have  coupling strength $\gamma_D$.
Asymmetric confinement of the 2DES, a SIA, yields  the  Rashba term which does not depend on the growth direction
\begin{equation}
-\frac{1}{2} \gamma{\bf B}_\kt{SO,R} = \alpha_2  ( \hat{e}_x p_y - \hat{e}_y p_x),
\end{equation}
 with $\alpha_2$ the  Rashba parameter.\cite{rashba_1,rashba_2}
Therefore the Cooperon Hamiltonian, in the case of Rashba and lin. and cubic Dresselhaus SOC is given by
\begin{equation}\label{H001}
 H_c:=\frac{\hat C^{-1}}{D_e}=({\bf Q}  +2 e {\bf A}_{\mathbf{S}} )^2+(m_e^2\epsilon_F\gamma_D)^2(S_x^2+S_y^2),
\end{equation}
with the effective vector potential
\begin{equation}
{\bf A}_{\mathbf{S}} =  \frac{m_e}{e}\hat\alpha{\bf S}= \frac{m_e}{e}\left(
\begin{array}{ccc}
 -\tilde\alpha_1 & -\alpha_2 & 0 \\
 \alpha_2 & \tilde\alpha_1 & 0
\end{array}
\right)
\left(\begin{array}{c}
S_x\\
S_y\\
S_z
\end{array}\right),
\end{equation}
with $\tilde\alpha_1=\alpha_1-m_e\gamma_D\epsilon_F/2$.\\
It can be easily shown that the Hamiltonian Eq.\,(\ref{H001}) has only non vanishing eigenvalues due to $(m_e^2\epsilon_F\gamma_D)^2$ in the 2D case.\\
The term with $(S_x^2+S_y^2)$, which is due to cubic Dresselhaus SOC, is not reduced by reason of the boundary in the diffusive case. However two triplet eigenvalues of this term depend on the wire width,
\begin{align}
 E_{QD1}={}&\frac{q_{s3}^2}{2},\\
 E_{QD2,3}={}&\frac{q_{s3}^2}{2}\left(\frac{3}{2}\pm\frac{\sin(Q_\kt{SO}W)}{2Q_\kt{SO}W}\right),
\end{align}
with $q_{s3}^2/2=(m_e^2\epsilon_F\gamma_D)^2$. In the following we are going to diagonalize the whole Hamiltonian and change the direction of the wire in the (001) plane.
\section{Spin Relaxation anisotropy in the (001) system}\label{sec:001grothDir}
\subsection{2D system}\label{sec:001grothDir2D}
We rotate the system in-plane through the angle $\theta$ (the angle $\theta=\pi/4$ is equivalent to [110]). This does not effect the Rashba term but changes the Dresselhaus one to\cite{Cheng2007,Wu2010}
\begin{align}
{}&\frac{1}{\gamma_D}H_{D[001]}=\nonumber\\
{}&\sigma_y k_y\cos(2\theta)(\langle k_z^2 \rangle-k_x^2)-\sigma_x k_x\cos(2\theta)(\langle k_z^2 \rangle-k_y^2)\nonumber\\
{}&-\sigma_y k_x\frac{1}{2}\sin(2\theta)(k_x^2-k_y^2-2\langle k_z^2 \rangle)\nonumber\\
{}&+\sigma_x k_y\frac{1}{2}\sin(2\theta)(k_x^2-k_y^2+2\langle k_z^2 \rangle),
\end{align}
with the wave vectors $k_i$.
The resulting Cooperon Hamiltonian, including Rashba and Dresselhaus SOC, reads then
\begin{align}\label{HrotatedWithDresselhaus}
 H_c={}&(Q_x+\alpha_{x1}S_x+(\alpha_{x2}-q_2)S_y)^2\nonumber\\
{}&+(Q_y+(\alpha_{x2}+q_2)S_x-\alpha_{x1}S_y)^2\nonumber\\
{}&+\frac{q_{s3}^2}{2}(S_x^2+S_y^2),
\end{align}
where we set
\begin{align}
 \frac{q_{s3}^2}{2}= {}&\left(m_e^2\epsilon_F\gamma_D\right)^2,\\
 \alpha_{x1}={}&\frac{1}{2}m_e\gamma_D\cos(2\theta)((m_ev)^2-4\langle k_z^2 \rangle),\\
 \alpha_{x2}={}&-\frac{1}{2}m_e\gamma_D\sin(2\theta)((m_ev)^2-4\langle k_z^2 \rangle)\\
={}&\left(q_1-\sqrt{\frac{q_{s3}^2}{2}}\right)\sin(2\theta)\\
={}&2m_e\tilde\alpha_1 \sin(2\theta),
\end{align}
with $q_1=2m_e\alpha_1$, $q_2=2m_e\alpha_2$.
We see that the part of the Hamiltonian which cannot be written as a vector field and is due to cubic Dresselhaus SOC does not depend on the wire direction in the (001) plane.
\subsubsection{Special case: Only lin. Dresselhaus SOC equal to Rashba SOC}
As a special example for the 2D case we set $q_{s3}=0$ and $q_{1}=q_2$. To simplify the search for vanishing spin relaxation we go to polar coordinates. Applying free wave functions (with $k_x,k_y$) to $H_c$, Eq.\,(\ref{HrotatedWithDresselhaus}), we end up with (singlet part left out)
\begin{equation}
\frac{H_c}{q_2^2}=\left(
\begin{array}{ccc}
 2+Q^2 & f_{\theta\phi} & -2\I\exp(2\I\theta) \\
 & 4+Q^2 & f_{\theta\phi} \\
 \text{c.c.} &  & 2+Q^2
\end{array}
\right)
\end{equation}
with $k_x/q_2=Q\cos(\phi)$, $k_y/q_2=Q\sin(\phi)$ and
\begin{equation}
 f_{\theta\phi}=(\I-1)\sqrt{2}\exp(\I\theta)(\cos(\phi+\theta)-\sin(\phi+\theta))Q.
\end{equation}
Vanishing spin relaxation is found at $Q=0$ for arbitrary values of $\theta$ (the spin with vanishing spin relaxation is pointing along the [110] direction\cite{Schliemann2003}). Another solution is found at $Q=2$ with the condition $\theta+\phi=3\pi/4$, which is equivalent to the $[\overline{1}10]$ crystallographic direction.\cite{Cheng2007}
\subsection{Quasi-1D wire}\label{sec:001grothDirWire}
In the following we consider spin and charge conserving boundaries. Due to the SOC we have the following modified Neumann condition\cite{Wenk2010}
\begin{align} \label{bc}
\left(-\frac{\tau}{D_e}{\bf n}\cdot\langle{\bf v}_F [\gamma{\bf B}_\kt{SO}({\bf k})\cdot{\bf S}]\rangle-\I\partial_{\bf n}\right) C |_{\partial S} ={}& 0,\\
\intertext{where $\langle ... \rangle$ denotes the average over the direction of ${\bf v}_F$ and ${\bf k}$ which we rewrite for the rotated x-y system}
(-\I \partial_y + 2 e  {\bf (A_{S})}_y ) C\left( x, y = \pm
\frac{W}{2}\right) ={}& 0,\enspace\forall x,\quad\quad
\end{align}
where ${\bf n}$ is the unit vector normal to the boundary $\partial S$ and x is the coordinate along the wire.
In order to do a diagonalization taking only the zero-mode into account, we have to simplify the boundary condition. A transformation acting in the transverse  direction is needed according to Eq.\,(\ref{HrotatedWithDresselhaus}):
$\hat{C}\rightarrow \tilde{\hat{C}} = U_A^{\phantom{\dagger}} \hat{C}U_A^\dagger$, by using the transformation
\begin{equation}\
 U=\eins_4-\I\sin(q_s y)\frac{1}{q_s}A_y+(\cos(q_sy)-1)\frac{1}{q_s^2}A_y^2
\end{equation}
with $A_y=(\alpha_{x2}+q_2)S_x-\alpha_{x1}S_y$ and\newline $q_s=\sqrt{(\alpha_{x2}+q_2)^2+\alpha_{x1}^2}$.
\subsubsection{Spin relaxation}
\begin{figure}[htbp]
\begin{center}
        \includegraphics[width=\columnwidth]{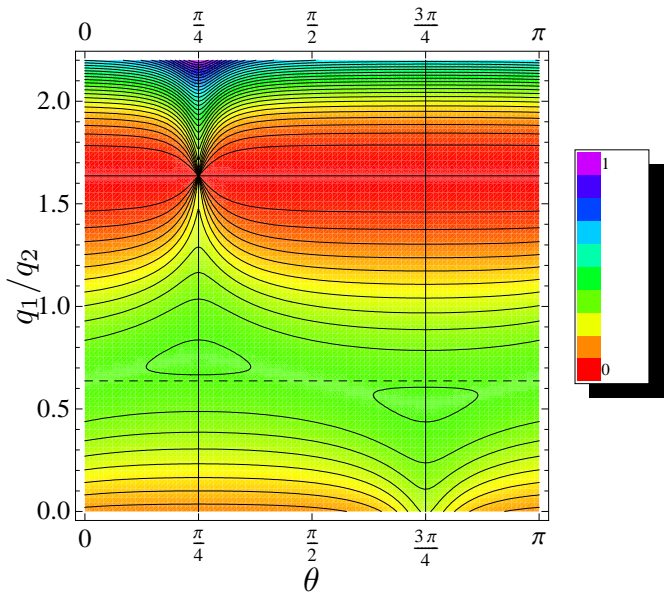}\\
        \caption{(Color online) Dependence of the $W^2$ coefficient in Eq.\,(\ref{E2min}) on the lateral rotation $(\theta)$. The absolute minimum is found for $\alpha_{x1}(\theta=0)=-q_2$ (here: $q_1/q_2=1.63$) and for different SO strength we find the minimum at $\theta=(1/4+n)\pi,\enspace n\in \mathbbm{Z}$ if $q_1<(qs3/\sqrt{2})$ (dashed line: $q_1=(qs3/\sqrt{2})$) and at $\theta=(3/4+n)\pi,\enspace n\in \mathbbm{Z}$ else. Here we set $q_{s3}=0.9$. The scaling is arbitrary.}\label{plot:theta}
\end{center}
\end{figure}

We diagonalize the Hamiltonian, Eq.\,(\ref{HrotatedWithDresselhaus}), after applying the transformation U, taking only the lowest mode into account. The spectrum of the Hamiltonian for small wire width, $W q_s<1$, is given by
\begin{align}
{}&E_{1/2}(k_x>0)=\nonumber\\
{}&k_x^2\pm k_x \left(2 q_{sm}-\frac{\left(\alpha_{x1}^2+\alpha_{x2}^2-q_2^2\right)^2}{12 q_{sm}}W^2 \right)\nonumber\\
{}&+\frac{3}{2}\frac{q_{s3}^2}{2}+q_{sm}^2\mp \frac{q_{s3}^2}{2k_x}\frac{\left(\alpha_{x1}^2+\alpha_{x2}^2-q_2^2\right)^2W^2}{96 q_{sm}}\nonumber\\
{}&-\frac{\left(\frac{q_{s3}^2}{2}+q_{sm}^2\right) \left(\alpha_{x1}^2+\alpha_{x2}^2-q_2^2\right)^2}{24 q_{sm}^2}W^2\,\label{E1/2},
\end{align}
\begin{align}
{}&E_{1}(k_x=0)=q_{s3}^2+q_{sm}^2\nonumber\\
{}&-\frac{(\alpha_{x1}^2+\alpha_{x2}^2-q_2^2)^2+\frac{q{s3}^2}{2}q_s^2}{12}W^2,\label{E1kx0}\\
{}&E_{2}(k_x=0)=\frac{q_{s3}^2}{2}+q_{sm}^2+\frac{q_{s3}^2}{2}\frac{q_2^2\alpha_{x1}^2}{3q_{sm}^2}W^2,\\
{}&E_3=k_x^2+\frac{q_{s3}^2}{2}+\frac{\left(\frac{q_{s3}^2}{2}+q_{sm}^2\right) \left(\alpha_{x1}^2+\alpha_{x2}^2-q_2^2\right)^2}{12 q_{sm}^2}W^2,
\end{align}
with $q_{sm}=\sqrt{(\alpha_{x2}-q_2)^2+\alpha_{x1}^2}$. First we notice that the only $\theta$ dependence is in the term $q_{sm}$, which disappears if the Dresselhaus SOC strength $\tilde\alpha_1$, which is shifted due to the cubic term, equals the Rashba SOC strength $\alpha_2$ and the angle of the boundary is $\theta=(1/4+n)\pi,\enspace n\in \mathbbm{Z}$.
Assuming the term proportional to $W^2/k_x$ to be small, the absolute minimum can be found at
\begin{align}
{}&E_{1/2,min}=\frac{3}{2}\frac{q_{s3}^2}{2}+\frac{\left(q_{sm}^2-\frac{q_{s3}^2}{2}\right) \left(\alpha_{x1}^2+\alpha_{x2}^2-q_2^2\right)^2}{24 q_{sm}^2}W^2\nonumber\\
\end{align}
which is independent of the width $W$ if $\alpha_{x1}(\theta=0)=-q_2$ and/or the direction of the wire is pointing in
\begin{equation}
 \theta=\frac{1}{2}\arcsin\left(\frac{2\langle k_z^2\rangle (m_e\gamma_D)^2((m_ev)^2-2\langle k_z^2\rangle)-q_2^2}{\left(m_e^3v^2\gamma_D-4\langle k_z^2\rangle m_e\gamma_D\right)q_2}\right).
\end{equation}
The second possible absolute minimum, which dominates for sufficient small width W and $q_{sm}\neq 0$ (compare with $E_2(k_x=0)$), is found at
\begin{align}\label{E2min}
{}&E_{3,min}=\frac{q_{s3}^2}{2}+\frac{\left(\frac{q_{s3}^2}{2}+q_{sm}^2\right) \left(\alpha_{x1}^2+\alpha_{x2}^2-q_2^2\right)^2}{12 q_{sm}^2}W^2.\nonumber\\
\end{align}
The minimal spin-relaxation rate is found by analyzing the prefactor of $W^2$ in Eq.\,(\ref{E2min}), Fig.\,(\ref{plot:theta}). We see immediately that in the case of vanishing cubic Dresselhaus or in the case where $\alpha_{x1}(\theta=0)=-q_2$ we have no direction dependence of the minimal spin relaxation. Notice the shift of the absolute minimum away from $q_1=q_2$ due to $q_{s3}\neq 0$. In the case of $q_1<(q_{s3}/\sqrt{2})$ we find the minimum at $\theta=(1/4+n)\pi,\enspace n\in \mathbbm{Z}$, else at $\theta=(3/4+n)\pi,\enspace n\in \mathbbm{Z}$, which is indicated by the dashed line in Fig.\,(\ref{plot:theta}).

\subsubsection{Spin dephasing}
\begin{figure}[htbp]
\begin{center}
        \includegraphics[width=\columnwidth]{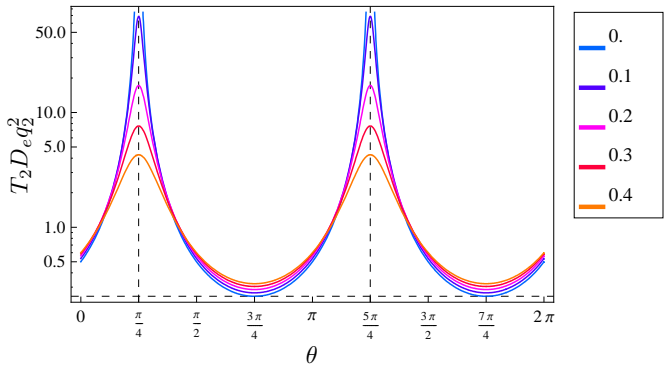}\\
        \caption{(Color online) The spin dephasing time $T_2$ of a spin initially oriented along the $[001]$ direction in units of $(D_eq_2^2)$ for the special case of equal Rashba and lin. Dresselhaus SOC. The different curves show different strength of cubic Dresselhaus in units of $q_{s3}/q_2$. In the case of finite cubic Dresselhaus SOC we set $W=0.4/q_2$. If $q_{s3}=0$: $T_2$ diverges at $\theta=(1/4+n)\pi,\enspace n\in \mathbbm{Z}$ (dashed vertical lines). The horizontal dashed line indicated the 2D spin dephasing time, $T_2=1/(4q_2^2D_e)$.}\label{plot:SpinDepWithMonteCarlo}
\end{center}
\end{figure}

Concerning spintronic devices it is interesting to know how an ensemble of spins initially oriented along the $[001]$ direction dephases in the wire of different orientation $\theta$. To do this analysis we only have to know that the eigenvector for the eigenvalue $E_1$ at $k_x=0$, Eq.\,(\ref{E1kx0}), is the triplet state $\ket{S=1;m=0} = ( \ket{\uparrow \downarrow} + \ket{\downarrow \uparrow})/\sqrt{2}  \equiv\linebreak \ket{\rightrightarrows}\hat=(0,1,0)^T$. This is equal to the z-component of the spin density whose evolution is described by the spin diffusion equation.\cite{Wenk2010} As an example we assume the case where cubic Dresselhaus term can be neglected and where the Rashba and lin. Dresselhaus SOC are equal. We notice that the dephasing is than width independent. At definite angles the dephasing time diverges - as for the in-plane polarized states with eigenvalue $E_2(k_x=0)$ - ,
\begin{equation}
 \frac{1}{\tau_s(W)}=2D_eq_2^2(1-\sin(2\theta))
\end{equation}
which is plotted in Fig.\,(\ref{plot:SpinDepWithMonteCarlo}). We have longest spin dephasing time at $\theta=(1/4+n)\pi,\enspace n\in \mathbbm{Z}$. For $\theta=(3/4+n)\pi,\enspace n\in \mathbbm{Z}$ we get the 2D result $T_2=1/(4q_2^2D_e)$, which is given by the eigenvalue of the spin relaxation tensor \cite{dyakonov72_2,dyakonov72_3,Wenk2010},
\begin{equation} \label{dp}
\frac{1}{\tau_{sij}} =  \tau\gamma^2\left( \langle  {\bf B}_{\rm SO}({\bf k})^2  \rangle \delta_{i j }-\langle { B}_{\rm SO}({\bf k})_i { B}_{\rm SO}({\bf k})_j\rangle\right)
\end{equation}
to the triplet state  $\ket{S=1;m=0}$.\\
This gives an analytical description of numerical calculation done by J.Liu \textit{et al.}, Ref.\,\onlinecite{Liu2009}.\\
Switching on cubic Dresselhaus SOC leads to finite spin dephasing time for all angles $\theta$. In addition $T_2$ is than width dependent. In the case of strong cubic Dresselhaus SOC where $q_{s3}^2/2=q_1^2=q_2^2$, the  dephasing time $T_2$ is angle independent and for $q_{s3}^2/2>q_1^2=q_2^2$ the minima in $T_2(\theta)$ change to maxima and vice versa.

\subsubsection{Special case: $\theta=0$}
In this case the longitudinal direction of the wire is [100].\newline
If we neglect the term proportional to $W^2/k_x$ in Eq.\,(\ref{E1/2}) the lowest spin relaxation is found to be
\begin{equation}
 \frac{1}{D_e\tau_s}=\frac{q_{s3}^2}{2}+\frac{\left(\alpha_{x1}^2-q_{2}^2\right)^2 \left(q_s^2+\frac{q_{s3}^2}{2}\right) W^2}{12 q_s^2}
\end{equation}
or
\begin{equation}
 \frac{1}{D_e\tau_s}=\frac{3 q_{s3}^2}{4}+\frac{\left(\alpha_{x1}^2-q_{2}^2\right)^2 \left(q_s^2-\frac{q_{s3}^2}{2}\right) W^2}{24 q_s^2},
\end{equation}
depending whether
\begin{equation}
 -\frac{q_{s3}^2}{4}+\frac{\left(\alpha_{x1}^2-q_{2}^2\right)^2 \left(q_s^2+3 \frac{q_{s3}^2}{2}\right) W^2}{24 q_s^2}
\end{equation}
is negative or positive. This shows that the cubic Dresselhaus term adds not only to the relaxation rate by a constant term but is also width dependent. However, this width dependence does not reduce the spin relaxation rate below $q_{s3}^2/2$.
\section{Spin relaxation in quasi-1D wire with [110] growth direction}\label{sec:110system}
To get the spin-relaxation in a $[110]$ QW with Rashba and Dresselhaus SOC again we have to rotate the spacial coordinate system of the Dresselhaus Hamiltonian  Eq.(\ref{Dresselhaus}) but now with the rotation matrix
\begin{equation}
R=\left(
\begin{array}{lll}
 \frac{1}{\sqrt{2}} & 0 & \frac{1}{\sqrt{2}} \\
 -\frac{1}{\sqrt{2}} & 0 & \frac{1}{\sqrt{2}} \\
 0 & 1 & 0
\end{array}
\right).
\end{equation}
We get
\begin{align}
 \frac{H_{D[110]}}{\gamma_D} ={}&\sigma_x(-k_x^2k_z-2k_y^2k_z+k_z^3)\nonumber\\
{}&+\sigma_y(4k_xk_yk_z)\nonumber\\
{}&+\sigma_z(k_x^3-2k_xk_y^2-k_xk_z^2).
\end{align}
The confinement in z-direction ($z\equiv$[110]) leads to $\langle k_z\rangle=\langle k_z^3\rangle=0$, and $\langle k_z^2\rangle=\int |\nabla\phi|^2dz$. The Hamiltonian for the QW in $[110]$ direction has then the following form\cite{Hassenkam1997}
\begin{equation}
 H_{[110]}=-\gamma_D\sigma_zk_x\left(\frac{1}{2}\langle k_z^2\rangle-\frac{1}{2}(k_x^2-2k_y^2)\right).
\end{equation}
Including the Rashba SOC ($q_2$), noting that its Hamiltonian does not depend on the orientation of the wire,\cite{Hassenkam1997} we end up with the following Cooperon Hamiltonian
\begin{equation}
 \frac{C^{-1}}{D_e}=\left(Q_x-\tilde q_1 S_z-q_2S_y \right)^2+(Q_y+q_2S_x)^2+ \frac{\tilde q_{3}^2}{2} S_z^2.
\end{equation}
with $\tilde q_1 = 2 m_e \frac{\gamma_D}{2} \langle k_z^2\rangle-\frac{\gamma_D}{2}\frac{m_e\epsilon_F}{2}$, $q_2 = 2 m_e \alpha_2$ and $\tilde q_3=(3m_e\epsilon_F^2(\gamma_D/2))$.
We see immediately that in the 2D case states polarized in the z-direction have vanishing spin relaxation as long as we have no Rashba SOC. Compared with the (001) system the constant term due to cubic Dresselhaus does not mix spin directions. Here we set the appropriate Neumann boundary condition as follows:
\begin{equation}
(-i\partial_y + 2 m_e \alpha_2 S_x) C\left(x,y = \pm\frac{W}{2}\right) =0,\enspace\forall x.\label{110boundaryCond}
\end{equation}
The presence of Rashba SOC adds a vector potential proportional to $S_x$.
Applying a non-abelian gauge transformation as before to simplify the boundary condition, we diagonalize the transformed Hamiltonian (App.\,(\ref{hamiltonian})) up to  second order in $q_2W$ in the 0-mode approximation.
\subsection{Special case: without cubic Dresselhaus SOC}\
The spectrum is found to be
\begin{align}
E_1={}&k_x^2+\frac{1}{12} \Delta^2 (q_2 W)^2,\label{110E1}\\
E_{2,3}={}&k_x^2+\frac{1}{24}\Delta^2\left(24-( q_2 W)^2\right)\nonumber\\
{}&\pm\frac{\Delta}{24}\sqrt{\Delta^2(q_2W)^4+4k_x^2(24-(q_2W)^2)^2},
\end{align}
with the lowest spin relaxation rate found at finite wave vectors $k_{x_\kt{min}}=\pm\frac{\Delta}{24}(24-(q_2 W)^2)$,
\begin{equation}
 \frac{1}{D_e\tau_s}=\frac{\Delta^2}{24}(q_2W)^2.
\end{equation}
We set $\Delta=\sqrt{\tilde q_1^2+q_2^2}$.
\subsection{With cubic Dresselhaus SOC}
If cubic Dresselhaus SOC cannot be neglected, the absolute minimum of spin relaxation can also shift to $k_{x_\kt{min}}=0$. This depends on the ratio of Rashba and lin. Dresselhaus SOC:\\
If $q_2/q_1\ll1$, we find the absolute minimum at $k_{x_\kt{min}}=0$,
\begin{align}
 E_{min1}={}&\frac{\tilde q_3+\tilde q_1^2+ q_2^2}{2}-\Delta_c+\frac{1}{12}\Delta_c(q_2 W)^2,
\end{align}
with
\begin{equation}
 \Delta_c=\frac{1}{2}\sqrt{(\tilde q_3+\tilde q_1^2)^2+2(\tilde q_1^2-\tilde q_3)q_2^2+q_2^4}.
\end{equation}
If $q_2/q_1\gg 1$, we find the absolute minimum at \newline $k_{x_\kt{min}}\approx\pm\frac{\Delta}{24}(24-(q_2 W)^2)$,
\begin{align}
E_{min2}= {}&k_{x_\kt{min}}^2-k_{x_\kt{min}}q_2\left(\frac{\tilde q_1^2}{q_2^2}+2\right)-\frac{\tilde q_3^2}{16 k_{x_\kt{min}} q_2}\nonumber\\
{}&+\tilde \Delta^2+\frac{\tilde q_3}{2}\left(\frac{\tilde q_1^2}{q_2^2}+1\right)\nonumber\\
{}&-\left(\frac{\tilde q_3 \tilde q_1^2}{12}-\frac{\tilde q_3^2 q_2}{3072 k_{x_\kt{min}}^3}-\frac{q_2^2}{24}(\tilde q_3-\tilde q_1^2)\right.\nonumber\\
{}&+\frac{ q_2^4}{24}-\left(\frac{\tilde q_1^2}{24}+\frac{q_2^2}{12}\right)q_2 k_{x_\kt{min}}\nonumber\\
{}&\left.-\frac{q_2}{k_{x_\kt{min}}}\left(\left(\frac{\tilde q_3^2}{128}+\frac{\tilde q_3 \tilde q_1^2}{192}\right)-\frac{\tilde q_3 q_2^2}{96}\right)\right)W^2.\label{110Emin2}
\end{align}
We can conclude that reducing wire width W will not cancel the contribution due to cubic Dresselhaus SOC to the spin relaxation rate.
\section{weak localization}\label{sec:WL}
In Ref.\,\citenum{Kettemann:PRL98:2007,Wenk2010} the crossover from WL to WAL due to change of wire width and SOC strength was explained in the case of a (001) system. Whether WL or WAL is present depends on the suppression of the triplet modes of the Cooperon. The suppression in turn is dominated by the absolute minimum of the spectrum of the Cooperon Hamiltonian $H_c$. The findings presented in Sec.\,\ref{sec:001grothDirWire} therefore point out that e.g. the crossover width, at which the system changes from WL to WAL, can shift with the wire direction $\theta$. Recently experimental results on WL/WAL by J. Nitta \textit{et al.}, Ref.\,\onlinecite{Nitta2010}, seem to show a strong dependence on growth direction.\\
In the (110) system the situation is different: In the 2D case it was shown by Pikus \textit{et al.}, Ref.\,\onlinecite{Hassenkam1997}, that in the absence of the Rashba terms the negative magnetoconductivity cannot be observed. In the case of a wire geometry we can conclude from Eqs.\,(\ref{110E1}-\ref{110Emin2}) that we have no width dependence if Rashba SOC vanishes. A change of the quantum correction to the static conductivity therefore cannot be achieved in this wire geometry by changing the wire width. The reason is the vector potential in the boundary condition, Eq.\,(\ref{110boundaryCond}), which only depends on the Rashba SOC.
\section{Diffusive-Ballistic Crossover}\label{sec:DiffBallis}
In the following we assume a (001) 2D system with both, Rashba and linear and cubic Dresselhaus SO coupling.\\
Experiments measuring WL in diffusive QW with SOC\cite{hu05_2,Schapers2009} are in great agreement with theoretical calculations by S. Kettemann, Ref.\,\citenum{Kettemann:PRL98:2007}. But considering e.g. the works Ref.\,\citenum{kunihashi:226601, Kallaher2010}, one realizes that the scope of application of the theory has to be extended to describe also the crossover to the ballistic regime, $l_e>W$. We have shown in Sec.\,\ref{sec:001grothDirWire} that the presence of cubic Dresselhaus SOC in the sample leads to a finite spin relaxation even for wire widths $Q_\kt{SO}W\ll 1$, regardless of the boundary direction in a (001) system.
To account for the ballistic case we have to modify the derivation of the Cooperon Hamiltonian, Eq.\,(\ref{H001}).
In the case of a wire where the mean free path $l_e$ is comparable to the wire width W, we cannot integrate in Eq.\,(\ref{Cooperon0}) over the Fermi surface in a continuous way. Instead, we assume $k_F/W$ to be finite and sum over the number of discrete channels $N=[k_F W/\pi]$, where $[\ldots]$ is the integer part. Because $H_{\gamma D}\sim \epsilon_F^2$ this constant term due to cubic Dresselhaus should reduce if we reduce the number of channels.
If we expand the Cooperon to second order in $( {\bf Q} +2 e {\bf A} + 2m_e \hat{a} {\bf S} )$ before averaging over the Fermi surface, $\langle\ldots\rangle$, and use the Matsubara trick, we get
\begin{align}\label{C1}
 \frac{C^{-1}}{D_e}={}&2f_1\left(Q_y+2\alpha_2S_x+2\left(\alpha_1-\gamma_D v^2\frac{f_3}{f_1}\right)S_y\right)^2\nonumber\\
{}& +2f_2\left(Q_x-2\alpha_2S_y-2\left(\alpha_1-\gamma_D v^2\frac{f_3}{f_2}\right)S_x\right)^2\nonumber\\
{}& +8\gamma_D^2v^4\left[\left(f_4-\frac{f_3^2}{f_2}\right)S_x^2+\left(f_5-\frac{f_3^2}{f_1}\right)S_y^2\right],
\end{align}
with $m_e=1$ and functions $f_i(\varphi)$ (App.\,\ref{Appen2}) which depend on the number of transverse modes N.
In the diffusive case we can perform the continuous sum over the angle $\varphi$ in Eq.\,(\ref{f1})-(\ref{f5}), and we receive the old result with $f_1=f_2=1/2$, $f_3=1/8$ and $f_4=f_5=1/16$:
\begin{align}
 H_c={}&(Q_y+2\alpha_2S_x+2\left(\alpha_1-\frac{1}{2}\gamma_D \epsilon_F\right)S_y)^2\nonumber\\
& + (Q_x-2\alpha_2S_y-2\left(\alpha_1-\frac{1}{2}\gamma_D \epsilon_F\right)S_x)^2\nonumber\\
& +(\gamma_D \epsilon_F)^2(S_x^2+S_y^2).
\end{align}
\subsection{Spin Relaxation at $Q_\kt{SO}W\ll 1$}
\begin{figure}[htbp]
\begin{center}
        \includegraphics[width=\columnwidth]{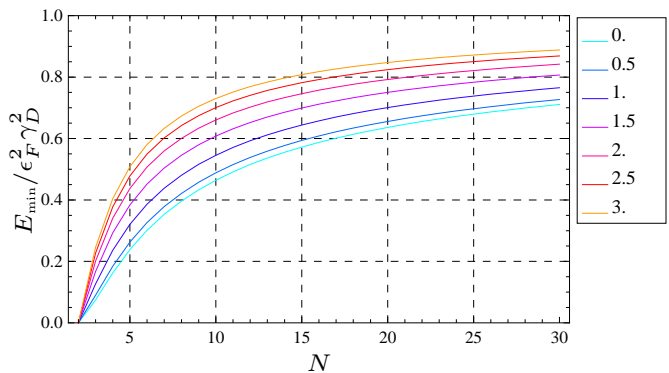}\\
        \caption{(Color online) The lowest eigenvalues of the confined Cooperon Hamiltonian Eq.\,(\ref{C1}), equivalent to the lowest spin relaxation rate, are shown for $Q=0$ for different number of modes $N=k_F W/\pi$. Different curves correspond to different values of $\alpha_2/q_s$.}\label{plot:modesCubicAlpha2}
\end{center}
\end{figure}

\begin{figure}[htbp]
\begin{center}
        \includegraphics[width=\columnwidth]{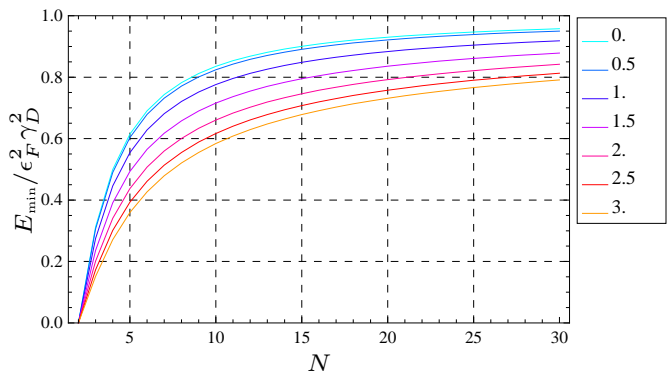}\\
        \caption{(Color online) The lowest eigenvalues of the confined Cooperon Hamiltonian Eq.\,(\ref{C1}), equivalent to the lowest spin relaxation rate, are shown for $Q=0$ for different number of modes $N=k_F W/\pi$. Different curves correspond to different values of $\alpha_1/q_s$.}\label{plot:modesCubicAlpha1}
\end{center}
\end{figure}

In the first section we analyzed the lowest spin relaxation in wires of different direction in a (001) system. We have shown, that for every direction there is still a finite spin relaxation at wire width which fulfill the condition $Q_\kt{SO}W\ll 1$ due to cubic Dresselhaus SOC. It is clear that this finite spin relaxation vanishes when the width is equal to the Fermi wave length $\lambda_F$. In the following we show how this finite spin relaxation depends on the number of transverse channels $N$. We show in Ref.\,\onlinecite{wenk3} that the findings are consistent with calculations going beyond the perturbative ansatz. This is possible in a similar manner as has been done previously in Ref.\,\onlinecite{km02_3}
for wires without SOC, which showed the crossover of the magnetic phase shifting rate, which had been known before in the diffusive and ballistic limit, only.
\\
To find the spectrum of the Cooperon Hamiltonian with boundary conditions as in Sec.\,(\ref{sec:001grothDirWire}), we stay in the 0-mode approximation in the Q space and proceed as before: According to Eq.\,(\ref{C1}), the non-Abelian gauge transformation for the transversal direction y is given by
\begin{equation}\label{trafo_ballistic}
 U=\exp\left(-\I\left[2\alpha_2 S_x+2\left(\alpha_1-\gamma_D v^2\frac{f_3}{f_1}\right)S_y\right] y\right).
\end{equation}
To concentrate on the constant width independent part of the spectrum we extract the absolute minimum at $Q=0$, Fig.\,(\ref{plot:modesCubicAlpha2}) and Fig.\,(\ref{plot:modesCubicAlpha1}). A clear reduction of the absolute minimum is visible. Due to the factor $f_3/f_1$ in the transformation U, the decrease of the minimal spin relaxation depends also on the ratio of Rashba and linear Dresselhaus SOC.

From Eq.\,(\ref{C1}) it is clear, that not only the $H_{\gamma_D}$ is affected by the reduction of the number of channels N but also the shift of the lin. Dresselhaus SOC, $\alpha_1$, in the orbital part. A model to extract the ratio of Rashba and lin. Dresselhaus SOC developed in Ref.\,\cite{epub7971} by Scheid \textit{et al.} did not show much difference between the strict 1D case and the non-diffusive case with wire of finite width. The results presented here should allow for extending the model to finite cubic Dresselhaus SOC. Deducing from our theory, the direction of the SO field should change with the number of channels due to the mentioned N dependent shift.
\section{Conclusions}
Summarizing the results, we have characterized the anisotropy and width dependence of spin relaxation in a (001) QW. There are special angles $\theta$ which are optimal for spin transport in quantum wires of finite width: The [110] and the $[\overline{1}10]$ direction. At [110] we find the the longest spin dephasing time $T_2$. If the absolute minimum of spin relaxation is found at [110] or $[\overline{1}10]$ direction depends on the strength of cubic Dresselhaus and wire width. The findings for the spin dephasing time are in agreement with numerical results. The analytical expression for $T_2$ allows to see directly the interplay between the cubic Dresselhaus SOC and the dimensional reduction, having effect on $T_2$. In addition we analyzed the special case of a (110) system and found the minimal spin relaxation rates depending on Rashba and lin. and cubic Dresselhaus SOC in the presence of boundaries. This results can be used to understand width and direction dependent WL measurements in QWs. Finally, we have shown how the reduction of channels in the wire reduces the finite spin relaxation rate which is due to cubic Dresselhaus SOC and does not reduce if the wire is small, $Wq_s\ll 1$, and diffusive, $W\gg l_e$. The change in channel number also changes the shift of lin. Dresselhaus SOC strength, $\tilde\alpha_1$. This has to be considered if extracting SOC strength from wires with only few transverse channels.
\begin{acknowledgments}
P.W. thanks the Asia Pacific Center for Theoretical Physics for hospitality. This research was supported by DFG-SFB508 B9 and
by WCU ( World Class University ) program through the Korea Science and Engineering Foundation funded by the Ministry of Education, Science and Technology (Project No. R31-2008-000-10059-0).
\end{acknowledgments}
\appendix
\section{Hamiltonian in [110] growth direction}\label{Appen1}
The Cooperon Hamiltonian in the 0-mode approximation is given as follows
\begin{equation}
 H_{c,0}=
\left(
\begin{array}{lll}
  A & B & C \\
  B^* & D & E \\
  C^* & E^* & F
\end{array}
\right)
+M_{q3},
\label{hamiltonian}
\end{equation}
 with
\begin{align}
 A={}&\frac{1}{4 q_2 W}(q_2 \left(4 k_x^2+3 \left(\tilde q_1^2+q_2^2\right)\right) W\nonumber\\
{}&-16 k_x \tilde q_1 \sin
   \left(\frac{q_2 W}{2}\right)+\left(\tilde q_1^2-q_2^2\right) \sin (q_2 W)),\\
B={}&\frac{\I \left(4 k_x \sin \left(\frac{q_2 W}{2}\right)-\tilde q_1 \sin (q_2 W)\right)}{\sqrt{2} W},\\
C={}&-\frac{q_2 \left(\tilde q_1^2+q_2^2\right) W+\left(q_2^2-\tilde q_1^2\right) \sin (q_2 W)}{4
   q_2 W},\\
D={}&\frac{q_2 \left(2 k_x^2+\tilde q_1^2+q_2^2\right) W+\left(q_2^2-\tilde q_1^2\right) \sin (q_2
   W)}{2 q_2 W},\\
E={}&\frac{\I \left(4 k_x \sin \left(\frac{q_2 W}{2}\right)+\tilde q_1 \sin (q_2 W)\right)}{\sqrt{2} W},\\
F={}&\frac{1}{4 q_2 W}(q_2 \left(4 k_x^2+3 \left(\tilde q_1^2+q_2^2\right)\right) W\nonumber\\
{}&+16 k_x \tilde q_1 \sin \left(\frac{q_2 W}{2}\right)+\left(\tilde q_1^2-q_2^2\right) \sin (q_2 W))
\end{align}
and the term due to cubic Dresselhaus SOC
\begin{align}
 {}& M_{q3}=\nonumber\\
 {}& q_3\left(
\begin{array}{lll}
  \frac{1}{4}\sinc(q_2 W)+\frac{3}{4} & 0 & \frac{1}{4}\sinc(q_2 W)-\frac{1}{4}\\
 0 & \frac{1}{2}-\frac{1}{2}\sinc(q_2 W) & 0 \\
 \frac{1}{4}\sinc(q_2 W)-\frac{1}{4} & 0 & \frac{1}{4}\sinc(q_2 W)+\frac{3}{4}
\end{array}
\right).
\end{align}
\section{Summation over the Fermi Surface}\label{Appen2}
The Cooperon Hamiltonian in the 2D case is given by
\begin{align}\label{C0}
H_c ={}& \tau v^2\{\langle \cos^2(\varphi)\rangle({\bf Q}+2m_e {\bf a.S})^2_x\nonumber\\
{}& +\langle \sin^2(\varphi)\rangle({\bf Q}+2m_e {\bf a.S})^2_y\nonumber\\
{}& +4m_e^2\gamma_D v^2\langle \cos^2(\varphi)\sin^2(\varphi)\rangle({\bf Q}+2m_e{\bf a.S})_x.S_x\nonumber\\
{}& -4m_e^2\gamma_D v^2\langle \sin^2(\varphi)\cos^2(\varphi)\rangle({\bf Q}+2m_e{\bf a.S})_y.S_y\nonumber\\
{}& +(2m_e^3\gamma_D v^2)^2(\langle \cos^2(\varphi)\sin^4(\varphi)\rangle S_x^2\nonumber\\
{}& +\langle \sin^2(\varphi)\cos^4(\varphi)\rangle S_y^2)\},
\end{align}
with wave vector $\mathbf Q$.
We set
\begin{align}
 m_e \equiv{}& 1,\\
f_1  :={}&\langle\sin^2(\varphi)\rangle\label{f1},\\
f_2  :={}&\langle\cos^2(\varphi)\rangle,\\
f_3  :={}&\langle\sin^2(\varphi)\cos^2(\varphi)\rangle,\\
f_4  :={}&\langle\sin^4(\varphi)\cos^2(\varphi)\rangle,\\
f_5  :={}&\langle\sin^2(\varphi)\cos^4(\varphi)\rangle\label{f5}.
\end{align}
Using the Matsubara trick we write
\begin{equation}
 \int_0^{2\pi}\frac{d\varphi}{2\pi}=\frac{2}{\pi N}\sum_{s=1}^N\frac{1}{\sqrt{1-\left(\frac{s}{N}\right)^2}}.
\end{equation}
This gives us
\begin{align}
f_1  ={}&\frac{2}{\pi N}\sum_{s=1}^{N-1}\frac{s^2}{N^2 \sqrt{1-\left(\frac{s}{N}\right)^2}},\\
f_2  ={}&\frac{2}{\pi N}\sum_{s=1}^{N}\sqrt{1-\left(\frac{s}{N}\right)^2},\\
f_3  ={}&\frac{2}{\pi N}\sum_{s=1}^{N}\left(\frac{s}{N}\right)^2 \sqrt{1-\left(\frac{s}{N}\right)^2},\\
f_4  ={}&\frac{2}{\pi N}\sum_{s=1}^{N}\left(\frac{s}{N}\right)^4 \sqrt{1-\left(\frac{s}{N}\right)^2},\\
f_5  ={}&\frac{2}{\pi N}\sum_{s=1}^{N}\left(\frac{s}{N}\right)^2 \left(1-\left(\frac{s}{N}\right)^2\right)^{\frac{3}{2}}.
\end{align}
Writing Eq.\,(\ref{C0}) in a compact way gives us Eq.\,(\ref{C1}).
\bibliographystyle{apsrev}
\bibliography{WenkKettemann2010.bib}
\end{document}